# Femtosecond pulse generation from a Topological Insulator mode-locked fiber laser

Hao Liu,[1] Xu-Wu Zheng,[1] Meng Liu,[1] Nian Zhao,[1] Ai-Ping Luo,[1] Zhi-Chao Luo,[1,*] Wen-Cheng Xu,[1,*] Han Zhang,[2] Chu-Jun Zhao,[3] and Shuang-Chun Wen[3]

[1]Laboratory of Nanophotonic Functional Materials and Devices, School of Information and Optoelectronic Science and Engineering, South China Normal University, Guangzhou, Guangdong 510006, China

[2]Key Laboratory of Optoelectronic Devices and Systems of Ministry of Education and Guangdong Province, Shenzhen University, 518060, China

[3]Key Laboratory for Micro-/Nano-Optoelectronic Devices of Ministry of Education, College of Physics and Microelectronic Science, Hunan University, Changsha 410082, China

[*]Corresponding authors: zcluo@scnu.edu.cn; xuwch@scnu.edu.cn

**Abstract:** We reported on the generation of femtosecond pulse in an anomalous-dispersion fiber ring laser by using a polyvinyl alcohol (PVA)-based Topological Insulator (TI), $Bi_2Se_3$ saturable absorber (SA). The PVA-TI composite has a low saturable optical intensity of 12 $MW/cm^2$ and a modulation depth of ~3.9%. By incorporating the fabricated PVA-TISA into a fiber laser, mode-locking operation could be achieved at a low pump threshold of 25 mW. After an optimization of the cavity parameters, optical pulse with ~660 fs centered at 1557.5 nm wavelength had been generated. The experimental results demonstrate that the PVA could be an excellent host material for fabricating high-performance TISA, and also indicate that the filmy PVA-TISA is indeed a good candidate for ultrafast saturable absorption device.

Ultrashort pulses have received much attention due to the great significance in a variety of applications, such as high-speed optical communications, spectroscopy, bio-medical imaging, and material processing [1-3]. As a simple and economic ultrashort pulse source, passively mode-locked fiber lasers have been extensively investigated in the past decade. To date, several passively mode-locked techniques, such as nonlinear polarization rotation (NPR) [4,5], nonlinear amplifying loop mirror (NALM) [6-8], and real saturable absorber (SA) [9-16], were employed to achieve ultrashort pulses in fiber lasers. Among them, inserting a high performance material based SA into the laser cavity was allowing successful mode-locking to be achieved with free cavity designs compared with other two techniques. Therefore, several types of SAs had been demonstrated to achieve the passive mode-locking operation of fiber laser. In recent years many researches had focused their attention on nano-material based SAs, which function as the rising candidates for ultrafast lasers due to their distinguished advantages in ultrafast recovery time, controllable modulation depth, easy fabrication and broadband saturable absorption [11-16]. In particular, graphene, a type of Dirac nanomaterials, had been successfully demonstrated as a medium that allows for high performance ultrafast fiber lasers [13-16].

Recently, Topological Insulator (TI), another type of Dirac nanoscale materials, was also proposed to as SA device to obtain ultrashort pulses in fiber lasers. F. Bernard et al. have firstly presented the saturable absorption behavior of TIs around 1550 nm [17]. In the following, by inserting the TIs-based SA into the laser cavity, Zhao et al. further demonstrated the mode-locked picosecond pulses in fiber laser [18,19]. Very recently, it was also shown that the TIs possess the saturable absorption around 1060 nm wavelength range, indicating that the TI is a material with broadband saturable absorption [20,21]. In addition, taking advantage of the large nonlinear refractive index of TI [22], we had obtained high repetition rate pulse with a microfiber-based TISA, demonstrating that the TIs could operate as both the high nonlinear photonic device and the SA in the laser system [23]. However, all the mode-locked fiber lasers based on TISA delivered the pulse-trains with the durations of a few picoseconds while the femtosecond pulse have not yet been achieved. Moreover, the TISAs for passive mode-locking reported before were fabricated by spraying TI on a quartz plate [18-20], and depositing it onto the fiber end facet [21,24] or microfiber [23]. In fact, for the purpose of flexibility, cost-effective and controllability, a filmy TISA composited by polyvinyl alcohol (PVA) or polymethyl methacrylate (PMMA) would be more favorable. As we know, the pulse characteristics, such as duration and stability, could be affected by the fabrication method of the nano-material-based SA [13-16]. Therefore, a question would arise as to whether the femtosecond pulse could be achieved by inserting a filmy TISA into a fiber laser.

In this work, we reported on the generation of femtosecond pulse from an anomalous-dispersion fiber ring laser by using a filmy TISA composited by the PVA. The measured Raman spectrum of the PVA-TISA shows that the TI: $Bi_2Se_3$ nano-sheets could maintain their structure when embedding into the PVA. The modulation depth and the saturable intensity are measured to be 3.9% and 12 MW/cm$^2$, respectively, which could make the fiber laser mode-locking at a low pumping threshold. By inserting the PVA-TISA into a fiber laser, ~ 660 fs output pulse centered at 1557.5 nm wavelength could be obtained. These results suggest that the TISA could be indeed a good candidate of ultrafast saturable absorption device, and also demonstrate that the PVA could be the excellent host for fabricating high-performance TISA.

The TI: $Bi_2Se_3$ nanosheets could be synthesized by various methods such as chemical vapor transport, mechanical exfoliations by Scotch tape or peeling by an atomic force microscope tip, molecular beam epitaxial growth etc . Here, the polyol method was utilized to synthesize $Bi_2Se_3$ nanosheets [25,26]. The dispersion enriched TI solution is prepared by ultra-sonicating $Bi_2Se_3$ nanosheets for 1 hour in acetone solution, as shown in Fig. 1(a). The concentration of TI/acetone solution is ~0.1 mg/ml. Then 1 mL TI/acetone solution is mixed with 5 mL aqueous solution of polyvinyl alcohol (PVA) and ultrasonicated for 30 minutes. The mixture is then evaporated at room temperature on a slide glass, resulting in the formation of a filmy PVA-TI composite, as presented in Fig. 1(b). Finally the prepared PVA-TI composite is placed between two fiber connectors to form a fiber-compatible SA, as shown in Fig. 1(c).

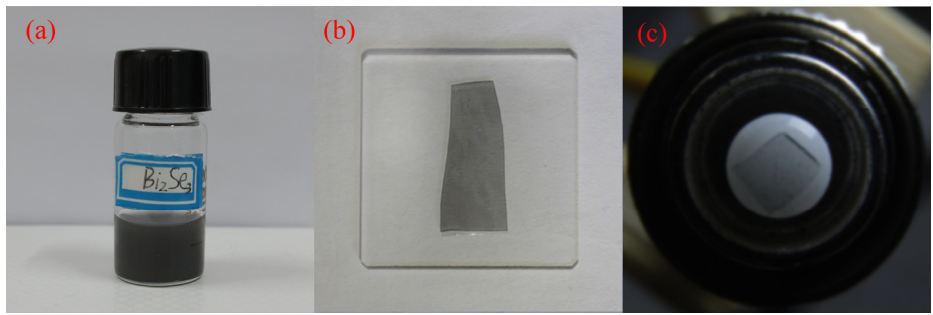

Fig. 1. (a) Image of $Bi_2Se_3$ acetone solution; (b) Image of TI-PVA film; (c) Image of the fabricated fiber-compatible PVA-TISA.

In order to check that whether the TI $Bi_2Se_3$ nano-sheets could maintain their structures or not after it has been embedded into PVA, we also performed the Raman spectrum analysis on the samples. Figure 2 compares the Raman spectra of the filmy PVA-TISA (blue curve) and pure PVA film (black curve) in the range of 0-1250 cm-1 using the 514 nm excitation line at room temperature by a Renishaw inVia micro-Raman system (Renishaw Inc., New Mills, UK). The spectrum of the TI-PVA composite, which shows the Raman peaks of both TI Bi$_2$Se$_3$ and PVA, can be seen as a superposition of PVA (black curve) and TI (red curve). Three typical Raman peaks of Bi$_2$Se$_3$ (Inset: Zoom-in view of the Raman peaks of Bi$_2$Se$_3$) centered at ~70 cm$^{-1}$, ~130 cm$^{-1}$ and ~173 cm$^{-1}$, were found in the Raman spectra of TI-PVA, which are correlated with the out-plane vibrational mode $A_{1g}^1$, in-plane vibrational mode $E_g^2$ and the out-plane vibrational mode $A_{1g}^2$ of Se-Bi-Se-Bi-Se lattice vibration, respectively [26,27]. Therefore, one can conclude that Bi$_2$Se$_3$ nanosheets could still keep their structures even though they are embedded into PVA matrix, also indicating that the optical property of TI does not encounter significant change.

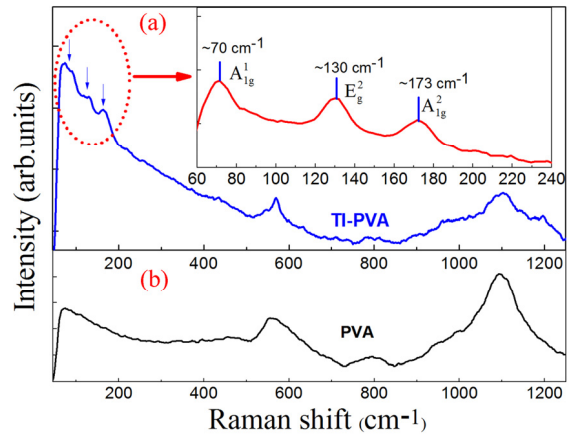

Fig. 2. Raman spectra of (a) TI-PVA, inset: zoom-in view of the Raman peaks of Bi$_2$Se$_3$; (b) PVA.

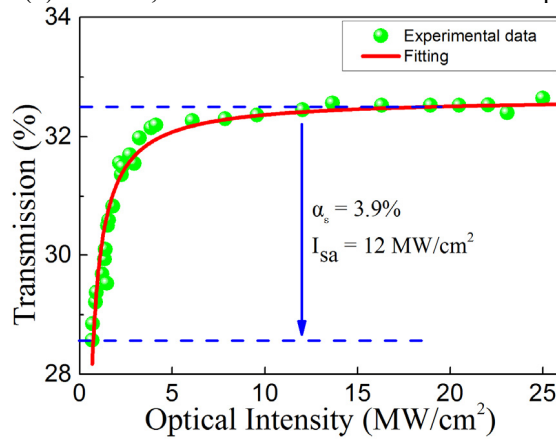

Fig. 3 Measured nonlinear saturable absorption curve and corresponding fitting curve.

For further investigating nonlinear optical characteristics of the fabricated PVA-TISA, the nonlinear absorption of TISA was measured by using a home-made femtosecond pulse pump source (center wavelength: 1554.4 nm; repetition rate: 26 MHz; pulse duration: ~500 fs). A commercial EDFA was used to amplify the femtosecond pulse to provide a wide range of pump power. A variable optical attenuator (VOA) was employed to control the input optical power of the TISA. Figure 3 provides the saturable absorption data of PVA-TISA and the corresponding fitting curve as a function of peak intensity. As can be seen here, the modulation depth is ~3.9% and the non-saturable loss is ~67.5%. It should be noted that the saturable intensity of the fabricated PVA-TISA is about two orders of magnitude less than the reported one (0.49 GW/cm$^2$) by depositing the TI onto a onto a thick quartz plate [19] and also smaller than the one (53 MW/cm$^2$) deposited onto fiber end facet [21], mainly benefiting from the few-layer structure. With such a low saturable intensity, it would be predicted that the threshold of mode-locking using the as-prepared PVA-TISA can be significantly reduced.

To check the laser performance by using the prepared PVA-TISA, the PVA-TISA was inserted into a fiber ring laser cavity. Figure 4 shows the schematic of the proposed fiber laser. A 4 m EDF with group velocity dispersion (GVD) parameter of D = -15 ps/nm/km was used as the gain medium fiber. The 976 nm pump light was coupled into the gain fiber by a wavelength division multiplexer (WDM). The polarization state of circulating light in the laser cavity could be controlled by adjusting a pair of polarization controllers (PCs). A polarization insensitive isolator (PI-ISO) was used to force the unidirectional operation of the laser cavity and a 10/90 coupler was used to output the laser emission. An optical spectrum analyzer (Anritsu MS9710C) and an oscilloscope (LeCroy, Wave Runner 104MXi, 1 GHz) with a photodetector (New Focus P818-BB-35F, 12.5GHz) were used to study the laser spectrum and output pulse train, respectively. In addition, the pulse duration was measured with a commercial autocorrelator (FR-103XL).

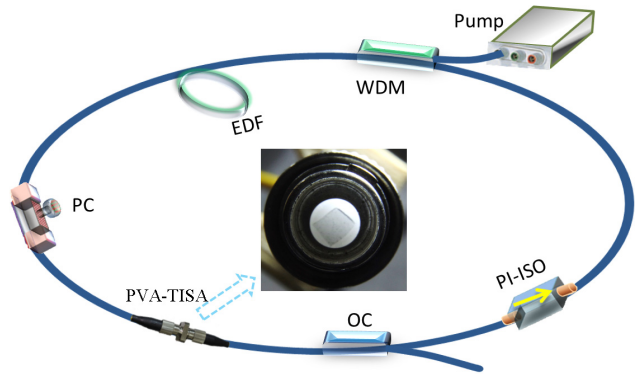

Fig. 4. Schematic of mode-locked fiber laser with a PVA-TISA.

Continuous wave operation started at a pump power of about 10 mW and the self-started mode-locking occurred at about 25 mW. For better performance of the fiber laser, the pump power was further increased to 90 mW. Figure 5 (a) shows the typical spectrum of the mode-locked pulses at 90 mW. Here, the central wavelength and the 3 dB spectral bandwidth are 1557.5 nm and 4.3 nm, respectively. The evident symmetric Kelly sidebands on the spectrum indicate that the mode locked laser is operating in the soliton regime. The corresponding pulse-train is presented in Fig. 5(b). The repetition rate is 12.5 MHz, which was determined by the 16.4 m cavity length. Then a commercial autocorrelator was employed to measure the pulse width. The measured result in Fig. 5(c) indicates that the fiber laser delivers a pulse-train with duration of 660 fs if the Sech2 intensity profile was assumed. Thus, the time-bandwidth product is about 0.351, which was slightly higher than 0.315, indicating that the mode-locked pulses are slightly chirped. In the experiment, only ~660 fs pulse was obtained by employing the prepared PVA-TISA. It should be noted that, apart from the quality of the SA, the pulse duration delivered from a fiber laser also could be affected by the cavity dispersion. Therefore, it would be expected that the shorter pulse duration can be obtained if the careful dispersion management technology is employed.

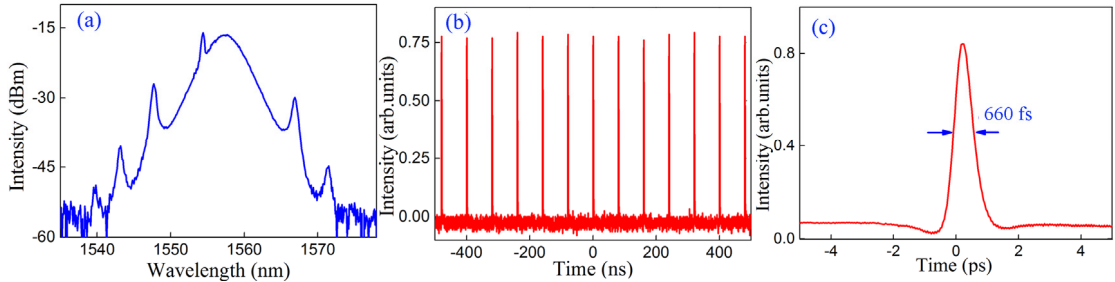

Fig. 5. Mode-locked operation. (a) Mode locked spectrum; (b) Corresponding pulse-train; (c) Corresponding autocorrelation trace.

To verify whether the mode locking operation was stable, the RF spectrum (Advantest R3131A) of the mode-locked pulse was measured, as presented in Fig. 6(a). The fundamental peak locates at the fundamental repetition rate of 12.5 MHz with a signal-to-noise ratio of over 55 dB, indicating that the stable pulse operation was obtained in this case. To further investigate the stability, we recorded the optical spectra of the mode-locked laser every 2-hour, as shown in Fig. 6(b). It should be noted that the central spectral peak locations, spectral bandwidth, spectral strength remained reasonably stable over the time period.

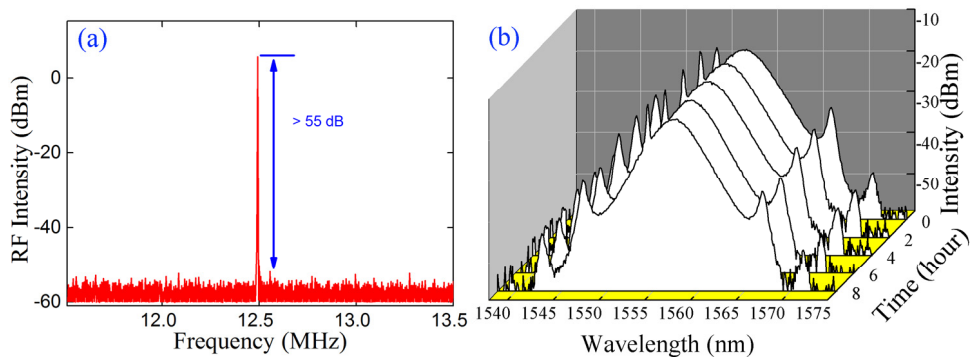

Fig. 6. Stability of femtosecond pulses. (a) RF spectrum; (b) Long term optical spectra measured at a 2-h interval.

In summary, we have demonstrated a femtosecond pulse fiber laser mode-locked by a filmy TISA. The filmy TISA, which is composited by the PVA, shows a modulation depth of

~3.9% and a low saturable optical intensity of 12 MW/cm$^2$. With the prepared PVA-TISA, the fiber laser delivers stable pulse-train with ~660 fs duration at the fundamental repetition rate of 12.5 MHz. Based on the experimental observations, it was expected that the PVA could be an excellent host material for fabricating high-performance TISA. The obtained results suggest that the PVA-TISA could be employed as a simple, low-cost ultrafast saturable absorption device for the applications such as mode-locked fiber lasers.

**Acknowledgments**

We would like to thank Yong-Fang Dong (South China Normal University) for help in measuring the Raman spectra. This work was supported in part by the National Natural Science Foundation of China (Grant No. 11074078, 61378036, 61307058, 11304101), the PhD Start-up Fund of Natural Science Foundation of Guangdong Province, China (Grant No. S2013040016320), and the Graduate Research and Innovation Foundation of South China Normal University, China (Grant No. 2013kyjj044).